\documentclass[12pt]{article}
\pdfoutput=1

\usepackage{amsmath,amssymb,amsbsy,amsfonts,latexsym,graphicx,hyperref}
\usepackage{color,array,subfigure}
\usepackage{cite}

\newcommand{\m}{\mu}

\newcommand{\nn}{\nonumber}

\newcommand{\tr}{{\rm tr}\,}

\allowdisplaybreaks

\evensidemargin \oddsidemargin

\def\m{{\mu}}

\def\nn{\nonumber}

\begin{document}


\begin{titlepage}

\renewcommand{\thefootnote}{\fnsymbol{footnote}}


\begin{flushright}
\end{flushright}

\baselineskip 7mm
\begin{center}
  {\Large \bf Thermodynamics of Inhomogeneously Mass-deformed ABJM Model and Pressure Anisotropy}
\end{center}

\begin{center}
\vskip 0.7 cm
  {Seungjoon Hyun$^{a}$\footnote{e-mail : sjhyun@yonsei.ac.kr}, Byoungjoon Ahn$^{a}$\footnote{e-mail : bjahn@yonsei.ac.kr},  Kyung Kiu Kim$^{b}$\footnote{e-mail : kimkyungkiu@sejong.ac.kr}, O-Kab Kwon$^{c}$\footnote{e-mail : okab@skku.edu} \\and Sang-A Park$^{a,d}$\footnote{e-mail : sapark@kias.re.kr}}

\vskip 0.5cm

{\it $^a\,$ Department of Physics, Yonsei University, Seoul 03722, Korea }\\
{\it $^b\,$ Department of Physics and Astronomy, Sejong University, Seoul 05006, Korea }\\
{\it $^c\,$ Department of Physics, Autonomous Institute of Natural Science, Institute of Basic Science,
Sungkyunkwan University, Suwon 16419, Korea } \\
{\it $^d\,$ School of Physics, Korea Institute for Advanced Study, Seoul 02455, Korea }

\end{center}

\thispagestyle{empty}

\vfill
\begin{center}
{\bf Abstract}
\end{center}
\noindent

In this paper we study the thermodynamics of black branes with a modulated complex scalar in the context of bulk and boundary theories. The modulation induces inhomogeneity to the dual field theory,  anisotropic pressure, and brane charge to the bulk geometry. The first law of thermodynamics and the Smarr relation are obtained using the off-shell ADT and the reduced action formalisms. We discuss the prescription for the mass of black branes, which relies on relevant and marginal deformations in the dual field theory. One of the cases is the gravity dual to a ABJM model with a sinusoidal mass function depending on a spatial coordinate. This is the first study of the deformed ABJM model at finite temperature including bulk thermodynamics.    
\\ [15mm]

\vspace{5mm}
\end{titlepage}

\baselineskip 6.6mm
\renewcommand{\thefootnote}{\arabic{footnote}}
\setcounter{footnote}{0}

\section{Introduction}
 
The asymptotically anti-de Sitter(AdS) black holes are very fascinating objects. They have much richer and more diverse structures than the asymptotically flat black holes. The topology of event horizon can be spherical, toroidal, and hyperbolic. This allows that the incorporation of scalar hairs is much easier and thus they can be readily used to probe the physics of the dual conformal field theory(CFT). By the AdS/CFT correspondence \cite{Maldacena:1997re,Witten:1998zw}, AdS black holes provide a powerful tool to study various strongly coupled CFTs including relevant or marginal deformations at finite temperature.

The motivation of our work comes from recently constructed supesymmetric models deformed by a spatially dependent sources \cite{Kim:2018qle,Kim:2019kns}, which are called Inhomogeneously mass-deformed ABJM (ImABJM) models. Those are based on the ABJM theory \cite{Aharony:2008ug} and the mass-deformed ABJM (mABJM) theory \cite{Hosomichi:2008jb,Gomis:2008vc}. As is well known, the ABJM theory describes N M2-branes probing an orbifold $\mathbb{C}^4/\mathbb{Z}_q$ and it is the $\mathcal{N}=6$ non-Abelian Chern-Simons (CS) matter theory with the gauge group $U_{-q}(N)\times U_{q}(N)$ at the CS level $q$. For lower q-levels, the supersymmetry of the theory is enhanced to $\mathcal{N}=8$ supersymmetry. The gravity duals of the ABJM theory is known as the 11-dimensional supergravity on $AdS_4\times S^7/\mathbb{Z}_q$. This ABJM theory is conformal and allows the supersymmetric mass deformation without loss of supersymmetry. Thus this deformation results in the non-conformal $\mathcal{N}=6$ mABJM theory.

The physical origin of the mass term is well understood in the M2 brane worldvolume picture \cite{Lambert:2009qw}. It turns out that the mass parameter $\hat{m}$ can be identified with the  four-form field strength $\mathcal{F}_{ABCD}$ of the 11-dimensional supergravity, which is turned on along the perpendicular directions to the worldvolume of the M2-branes. Here the indices $ABCD$ denote $\mathbb{C}^4$ directions. Thus the four-form field strength can be regarded as a scalar on the world volume. Suppose the four-form allows the dependence of a worldvolme coordinate $x$, then the mass parameter becomes a function $\hat{m}(x)$ and another components of the field strength $\mathcal{F}_{x\,A'B'C'} \sim \hat{m}'(x)$ should be turned on by the Bianchi identity \cite{Kim:2018qle}. Here the indices $A'B'C'$ again stand for some perpendicular directions. This is actually regarded as an one-form field strength on the worldvolume point of view. Thus this one-form field strength\footnote{In the bulk picture this one-form is equivalent to a massless scalar which would be explained in section 2.} induced by the inhomogeneity can be dualized to a two-form field strength. In the dual gravity picture, this becomes a three-form field strength including the radial direction and carries a brane charge. We show this structure by a field redefinition of the complex scalar field in the action (\ref{bulk_action}). Although we focused on the AdS Q-soliton with a complex scalar hair, a part of study on the black branes has been done recently without introduction of the brane charge in \cite{Ahn:2019pqy}.

In order to describe deformed ABJM theories in the strong coupling limit, we consider the corresponding gravity action. Since, as we mentioned, the vacuum of the ABJM theory is dual to $AdS_4\times S^7/\mathbb{Z}_q$, one can consider a consistent truncation of the $D=11$ supergravity on $S^7$ naturally. This truncation leads to the $\mathcal{N}=8$ $SO(8)$ gauged supergravity in 4-dimensions \cite{deWit:1981sst}. Also, a further truncation can be taken and the final resultant theory becomes $\mathcal{N}=4$ SO(4) gauged supergravity in 4-dimensions \cite{Das:1977pu}. Therefore the gravity duals to the ImABJM models can be expected to be solutions of an action which belongs to the $\mathcal{N}=4$ gauged supergravity and also the $SO(4)\times SO(4)$ invariant sector of $\mathcal{N}=8$ gauged supergravity.

In \cite{Gauntlett:2018vhk,Arav:2018njv}, the authors considered a special case of $\mathcal{N}=4$ supergravity action coupled to a single chiral multiplet. See (\ref{bulk_action}) for the bosonic part of the action. This action also belongs to $\mathcal{N}=1$ supergravity theory in 4-dimensions\footnote{This action can also be obtained from the $\mathcal{N}=2$ STU gauged supergravity \cite{Cvetic:1999xp,Cvetic:2000tb}.}. In those works, the authors conjectured that the gravity solutions they found are dual to the $\mathcal{N}=3$ ImABJM models~\cite{Kim:2018qle} with spatially modulated mass functions. In particular, the supersymmetric domain wall solution with the, so-called, Q-lattice ansatz  \cite{Donos:2013eha,Gauntlett:2018vhk} shows a novel boomerang RG flow. They also showed that the gravity solutions are consistent with Ward identities and the physical quantities in the ImABJM models. It deserves to note that the solutions 
in this 4-dimensional action can be uplifted to the D=11 supergravity theory. So it would be a promising direction to find the 11-dimensional origin of the gravity solutions.

On the other hand, it is also interesting to study the gravity duals of the ImABJM models at finite temperature. In this paper, we are devoted to this problem. In order to make the ploblem simpler, we focus on solutions of the bosonic action (\ref{bulk_action}) for the sinusoidal mass function. As a natural analysis of the system, we use the field redefinition (Equation (3) $z=\tanh\rho e^{i\chi} $ ) mentioned earlier. Then we can introduce the dualized three-form field strength from the phase of the complex scalar. A nontrivial three-form carrying the brane charge renders the thermodynamics of the black branes much richer than the vanishing charge configuration corresponding to a homogeneous system. We complete the thermodynamics of the black brane with the complex scalar hair, or equivalently the real scalar and the three-form field strength, describing relevant and marginal deformations.

Also, the inhomogeneity considered in the deformations induces anisotropic structure of the thermodynamics, even though we introduce a little bit simple form of the inhomogeneity, Q-lattice ansatz. Interestingly, the bulk pressure\footnote{The bulk pressure was introduced by the cosmological constant to extend the bulk black hole thermodynamics \cite{Kubiznak:2016qmn}.} and the charge encode the inhomogeneity successfully. In addition we identify the pressure with the bulk pressure at the horizon and this prescription leads to a consistent thermodynamic volume. This is the first work of the hairy bulk pressure with numerically generated black brane solutions.

The black hole thermodynamics is completed by the first law and the Smarr relation \cite{Smarr:1972kt}. We use the off-shell Abbott-Deser-Tekin(ADT) formalism invented and developed in \cite{Abbott:1981ff,Deser:2002rt,Kim:2013zha,Hyun:2014sha,Hyun:2014nma,Peng:2016qnz} to find the first law. In addition the reduced action formalism \cite{Banados:2005hm,Hyun:2015tia,Ahn:2015shg} is considered to obtain the Smarr relation. We provide the thermodynamic first law not only for the bulk system but also for the boundary field theory, which, in general, is not easy to achieve with numerical solutions. The Smarr relation relates data of the boundary of AdS space to data at the horizon. In the bulk side, it gives the consistent relation between the horizon quantity, such as the Bekenstein-Hawking entropy, the Hawking temperature, and the pressure at the horizon, and the physical quantities evaluated at the boundary, the black hole mass and the brane charge carried by the three-form field strength. In the dual field theory, the Smarr relation gives us correct expressions of the thermodynamic potentials. In particular, we analyze the thermodynamics of the Q-lattice black brane which is dual to a $\mathcal{N}=3$ ImABJM model with a relevant deformation related to mass function, $\hat{m}(x)= \hat{m}_0 \sin k x$. This is the first study of the ImABJM model at finite temperature accompanied by our preliminary study for the Q-lattice black brane \cite{Ahn:2019pqy}.

This paper is organized as follows. In section 2 we define a field redefinition which makes it easy to introduce the three-form field strength. In addtion we provide various physical quantities and find numerical solutions. In section 3 we summarize the off-shell ADT formalism and the reduced action formalism and find the first law and the Smarr relation of the black brane solutions. In section 4 we focus on the neutral black brane case with the vanishing modulation or brane charge and we discuss the prescription about determination of the black hole mass for marginal and relevant deformations. In section 5 we reveal the thermodynamics of the charged black branes in terms of the bulk and the boundary interpretations. In section 6 we investigate the thermodynamics of the $\mathcal{N}=3$ ImABJM model with the sinusoidal mass function. In section 7 we summarize our work and provide future directions.

\section{Black Brane Solutions}\label{BBG}

In this section we present solutions of the action included in SO(4)$\times$SO(4) invariant sector of the $\mathcal{N}=8$ gaugeed supergravity in 4-dimesions. The solutions have scalar hair describing the two-form gauge field charge. The neutral solution corresponds to the marginal deformation in the dual conformal field theory(CFT) \cite{Hertog:2004dr}. The other charged solution, which was recently found in \cite{Ahn:2019pqy}, is dual to a mass-deformed ABJM theory which is generated by spatially dependent source functions. It can also be regarded as dual to the 3-dimensional CFT in the background of non-trivial charge distribution and corresponding chemical potential.
\subsection{The supergravity action and the brane configuration}
 Our starting action is  the bosonic part of  the supergravity action which is given by\footnote{We use the conventions: $8\pi G=1,\, l=1$.}
\begin{equation}\label{bulk_action}
S_B=\frac{1}{2}\int d^4x \sqrt{-g} \big( R - \frac{2}{(1-|z|^2)^2}\partial_\m z \, \partial^\m \bar{z} +\frac{2(3-|z|^2)}{1-|z|^2}\big)\,.
\end{equation}
The vacuum solution  is the 4-dimensional AdS spacetime with the cosmological constant
$\Lambda=-3$. 
The mass of the complex scalar field $z=X+iY$ is given by  $m^2=-2$ and thus in the hairy solutions the scalar field can have two normalizable complex modes  at the boundary, one of which is interpreted as the source of the dual boundary operator  and the other its expectation value.
In \cite{Ahn:2019pqy}, the numerical  black brane solutions were found in which 
the  complex scalar field $z$ 
is found to be in the Q-lattice configuration as
\begin{align}
z(r,x) = |z|e^{ikx} \,,
\end{align}
 to see the modulation effect in the dual mass-deformed ABJM model. In this paper we give an alternative  form of the action which is connected with the action (\ref{bulk_action}) by the field redefinition and use both action to describe the thermodynamics of the bulk gravity as well as  of the boundary CFT.

The complex scalar field $z$ can be decomposed as the modulus and phase fields as
\begin{equation}
z= \tanh\rho\, e^{i\chi}~.
\end{equation}
The action (\ref{bulk_action}) can be rewritten in terms of these fields as
\begin{equation}\label{bulk_action2}
S_B=\frac{1}{2}\int d^4x \sqrt{-g} \left( R - 2(\partial\rho)^{2}- \frac{1}{2}\sinh^{2}(2\rho)(\partial\chi)^{2} +2(2+\cosh (2\rho))\right)~.
\end{equation}
There is no potential term for the phase field and thus the equation of motion for the scalar field $\chi$,
\begin{equation}
\nabla_{\m}(\sinh^{2}(2\rho)\,\partial^{\m}\chi)=0~,
\end{equation}
suggests that we can introduce a two-form gauge field $C_{\mu\nu}$ whose field strength is Poincar\'e dual  to $\chi$ as
 \begin{equation}\label{dC1}
F_{(3)}\equiv dC=\sinh^{2}(2\rho) \ast d\chi\,.
\end{equation}
In terms of this two-form gauge field, the alternative form of the bulk action can be written as 
\begin{equation}\label{bulk_action3}
S=\frac{1}{2}\int d^4x \sqrt{-g} \left( R - 2(\partial\rho)^{2}- \frac{1}{2}\sinh^{-2}(2\rho)\,|dC|^{2} +2(3+ 2\sinh^{2} \rho)\right)\,.
\end{equation}
One may note that the two-form gauge field is effectively decoupled at the AdS vacuum as the scalar field $\rho$ vanishes and so does the effective gauge coupling.

\subsection{The neutral black brane solution with scalar hair}
In this subsection, we consider the neutral black brane solution with a real scalar hair. The metric ansatz of the black brane solution with planar symmetry can be taken as
\begin{align}
&ds^2 = -U(r) e^{2W_0(r) } dt^2 + \frac{ dr^2}{U(r)}+r^{2} (dx^2 +dy^2)
\end{align}
with non-vanishing scalar field $\rho(r)$.
The regularity condition for the scalar field at the horizon $(r=r_h)$ is given by
\begin{align}
&\rho '(r_h)=-\frac{2\sinh(2\rho_h)}{r_{h}(2+\cosh(2\rho_h))}~,
\end{align}
where $ \rho '(r)$ denotes the derivative of the scalar field $\rho(r)$ with respect to the coordinate $r$.
The asymptotic expansions of the metric and the scalar field are given by
\begin{align}
U(r) \sim~ & r^{2}+\rho_1^2-\frac{m}{r}+\frac{2\rho_2^2}{r^2}- \frac{m \rho_1^2 }{6 r^3} + \cdots~, \nonumber\\
W_0(r) \sim~ & -\frac{\rho_1^2}{2 r^2}- \frac{4\rho_1 \rho_2}{3r^3}+ \cdots~, \nonumber \\
\rho( r)\sim~ & \frac{\rho_1}{r}+\frac{\rho_2}{r^2}-\frac{\rho_1^3 }{6r^3} +\cdots.
\end{align}
The temperature and the entropy density of this black brane are determined as
\begin{align} 
T =\frac{1}{4\pi} e^{W_0(r_h)}U'(r_h)
 = \frac{r_h}{4\pi }e^{w_0}\Big(2+\cosh(2\rho_h)\Big)\,, \qquad
s = 2\pi r_h^2\,,
\end{align}
where the second equality in the expression of the temperature $T$ comes from the equation of motion.
We choose  that the fields, $\rho(r)$ and $W_{0}$,  vanish as they approach the asymptotic boundary, and introduce the values of those fields at the event horizon as $\rho_{h}= \rho(r_{h})$ and $w_{0}=W_{0}(r_{h})$. This solution was given and analyzed in \cite{Hertog:2004dr}. 

\subsection{The charged black brane solution with scalar hair}
Now we present the charged black brane solution dual to the massive deformation of ABJM model with spatially dependent mass.
We turn on the  phase scalar field as $\chi=kx$. In the dual two-form field perspective, the quantum number $k$ corresponds to the charge of the two-form gauge field. Therefore the configuration is that the charged branes spanning in the $y$-direction are located  periodically in the $x$-direction. As alluded in the introduction, 
from the perspective of 11-dimensional supergravity on $AdS_4\times S^7/\mathbb{Z}_q$, this arises as the turning on the non-trivial four-form field strength. This would correspond to  the insertion of  
 the additional $k$ M5-brane wrapping 4-cycles in  $S^7/\mathbb{Z}_q$ and stretching in the $y$-direction.

\begin{figure}[ht!]
\centering
    \subfigure[ ]
    {\includegraphics[width=6.5cm]{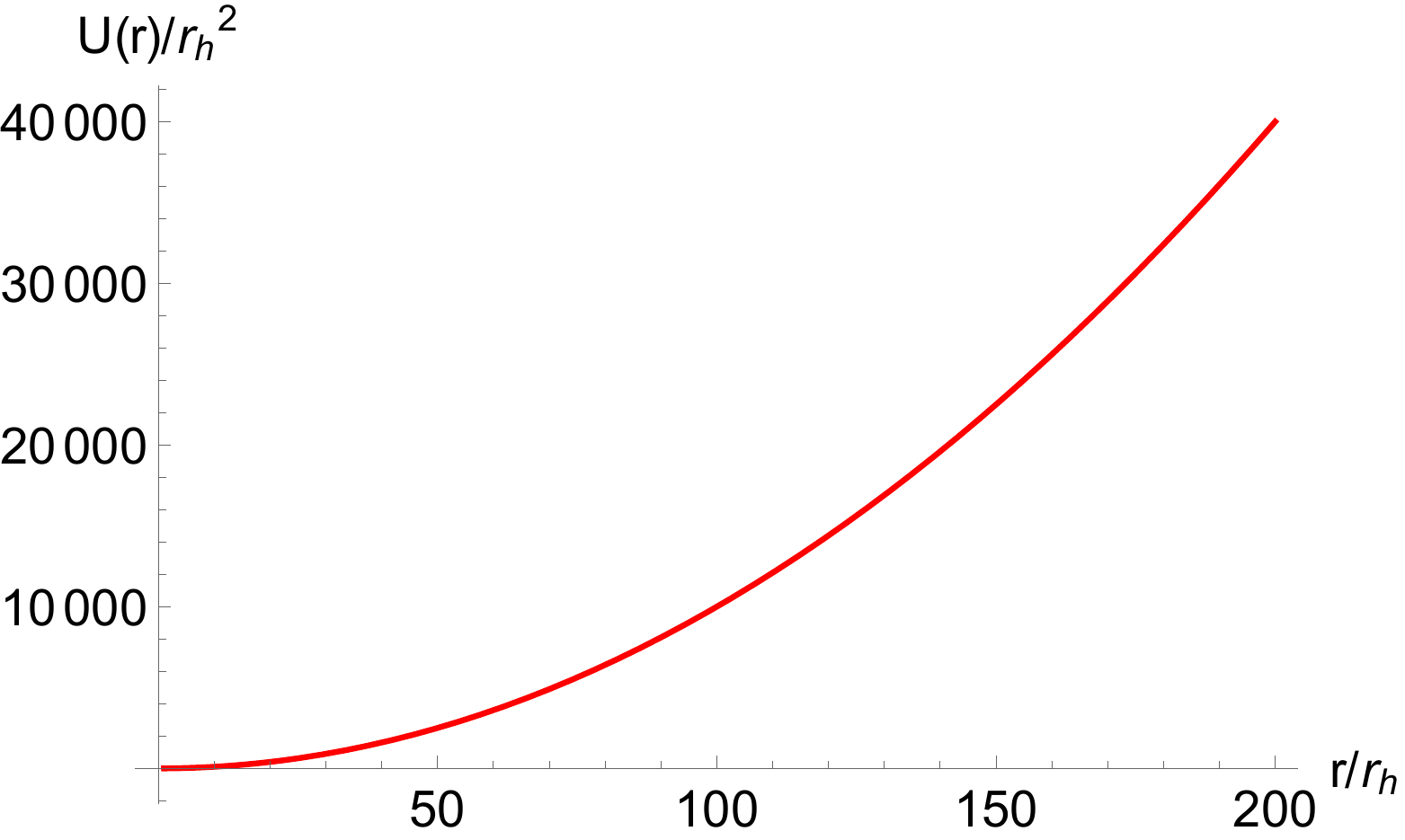}  }
\subfigure[ ]
{\includegraphics[width=6.5cm]{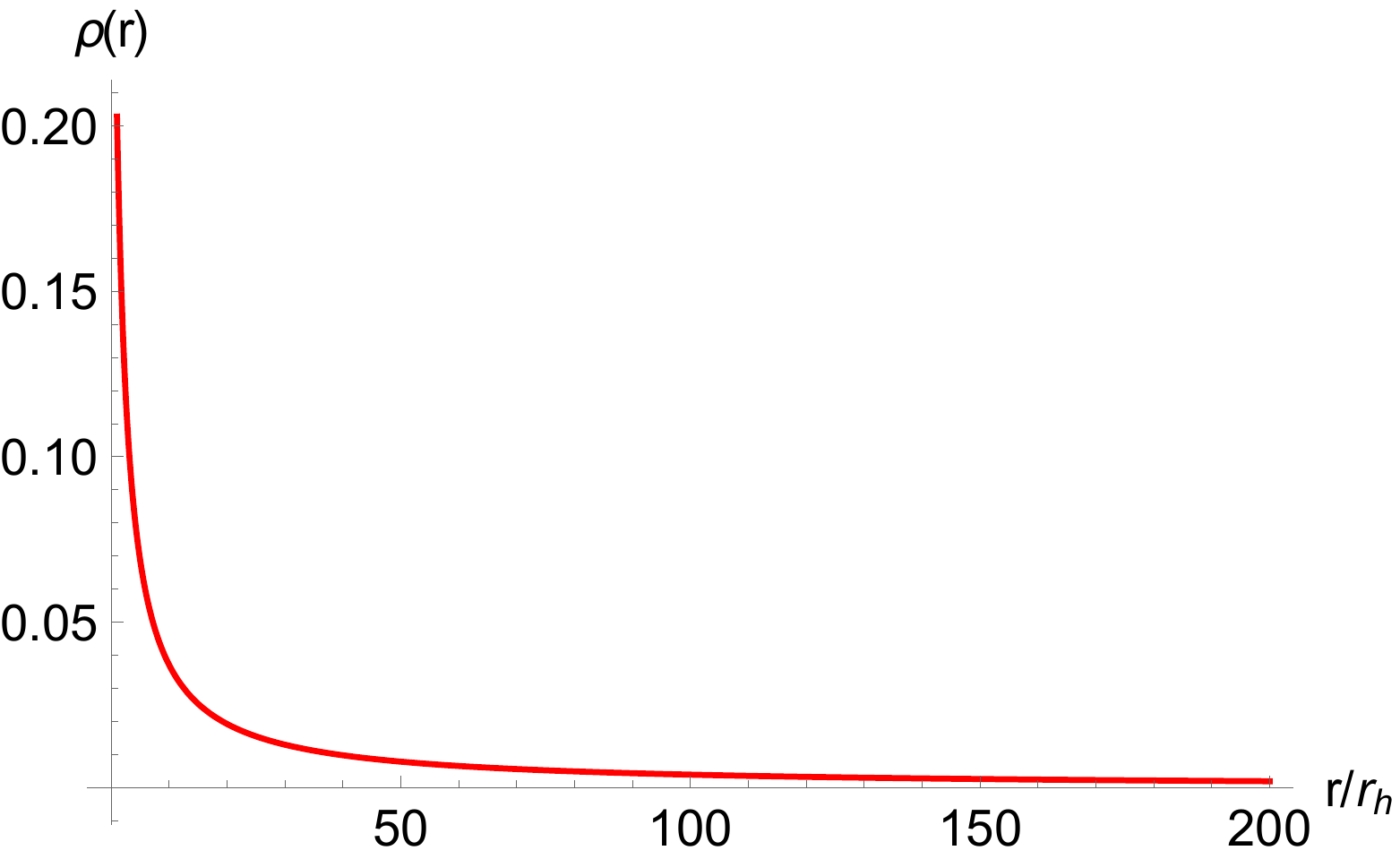}  }\\
    \subfigure[ ]
    {\includegraphics[width=6.5cm]{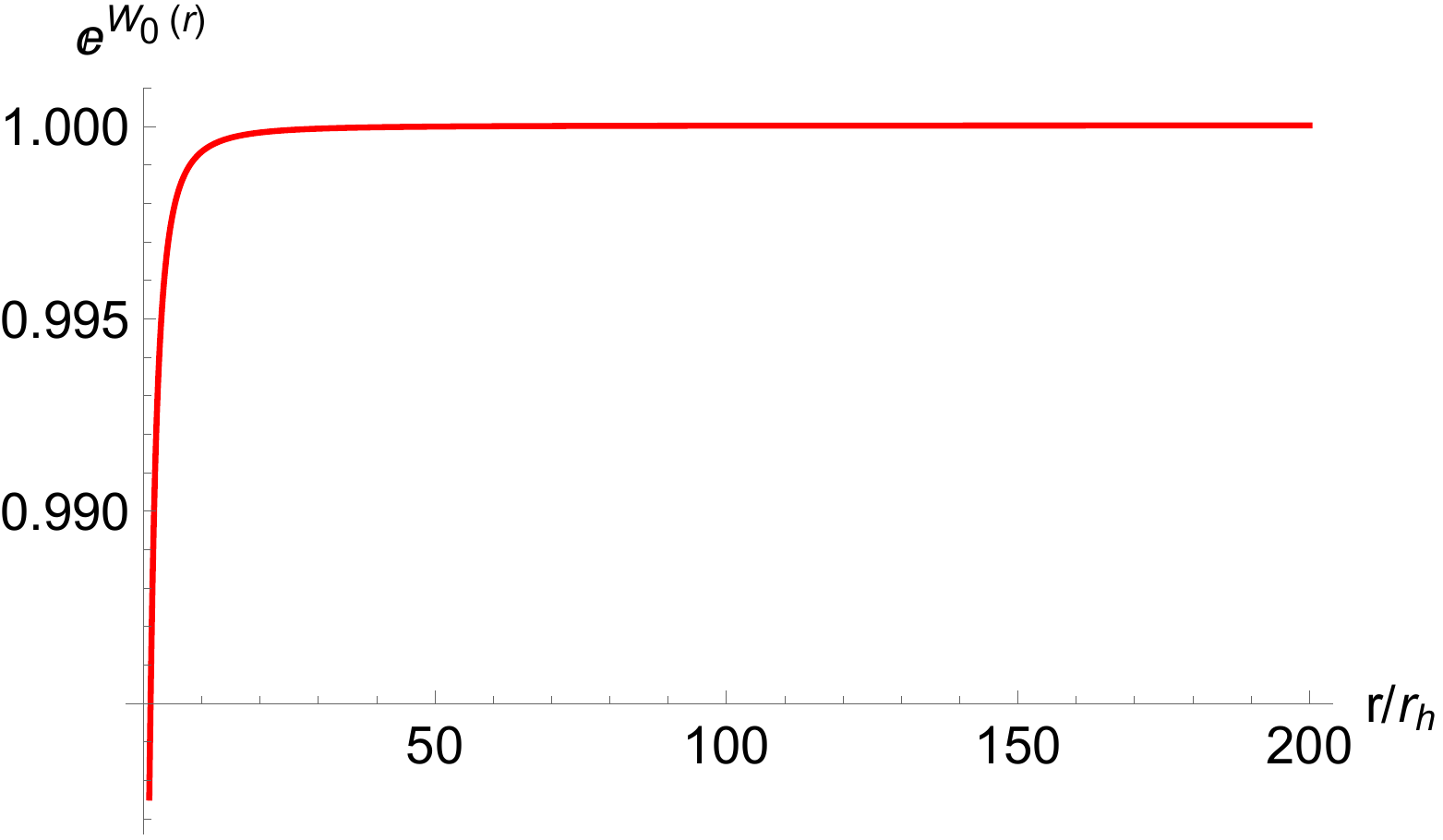}  }
\subfigure[ ]
{\includegraphics[width=6.5cm]{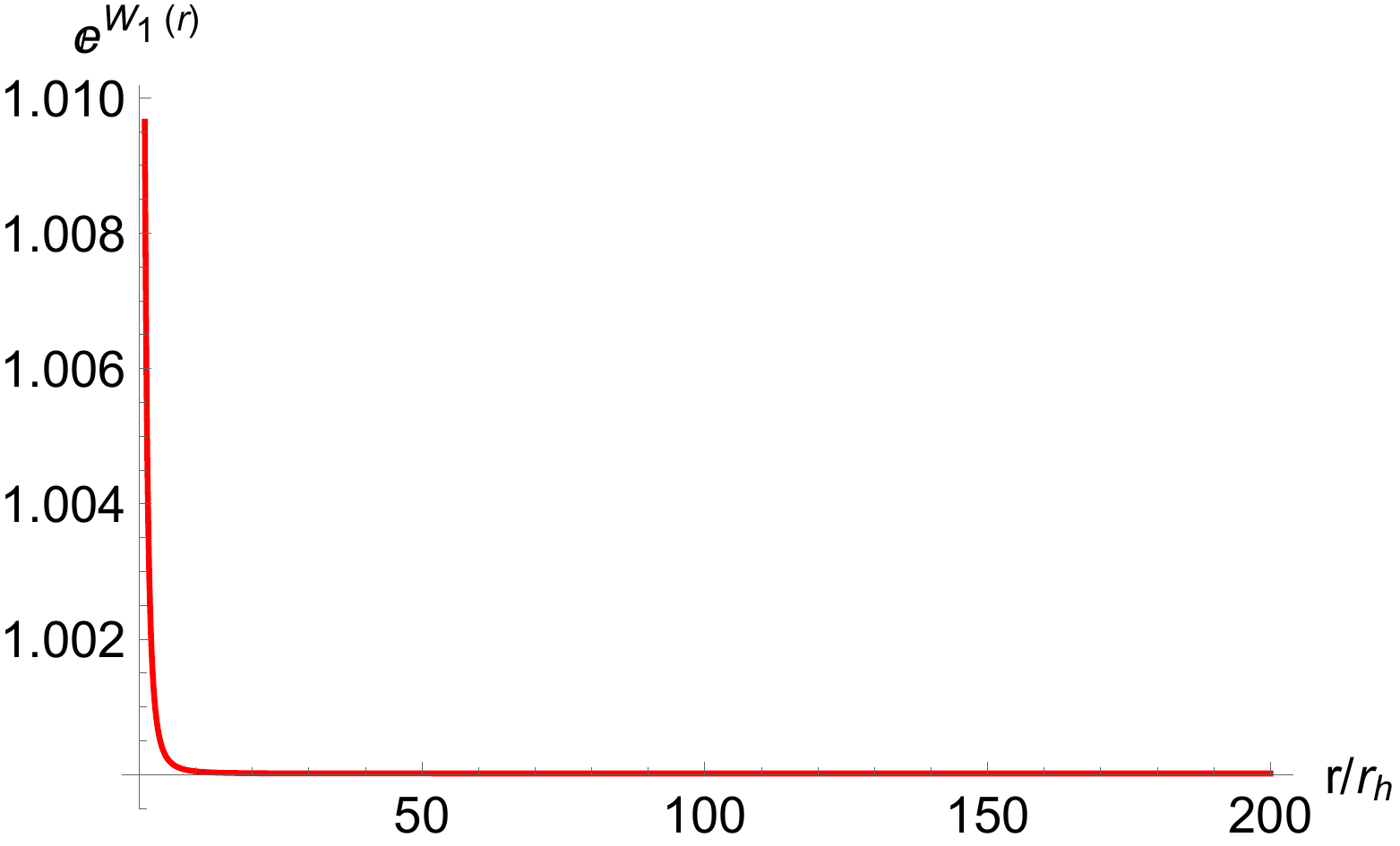}  }

    \caption{A numerical solution of the black branes with $k=0.8 \, r_{h}$, $\rho_{h}=0.2$, $w_0=-0.0176$ and $w_1=0.0096$
} \label{fig:HBH01}
\end{figure}

The ansatz for the metric is taken as
\begin{align}\label{BBrane00} 
&ds^2 = -U(r) e^{2W_0(r) } dt^2 + \frac{ dr^2}{U(r)} +r^{2} ( e^{2W_1(r) } dx^2+ dy^2) \,,
\end{align}
where we need to add the metric function $W_1(r)$ to incorporate the anisotropy due to the non-vanishing two-form gauge field. 
The regularity condition for the scalar field at the horizon is given by
\begin{align}
&\rho '(r_h)=-\frac{ \sinh (2 \rho_h)(2 r_h^2- e^{-2 w_{1}} k^2  \cosh (2 \rho_h) )}{2 r_h^3 (2+\cosh (2 \rho_h))}\,,
\end{align}
where the metric function $W_{1}$ is chosen to vanish at the asymptotic boundary by taking an appropriate value $w_{1}=W_{1}(r_{h})$ at the event horizon.
The asymptotic behavior of the fields is found to be
\begin{align}\label{AsympExp}
U(r) \sim~ & r^2+\rho_1^2-\frac{m}{r}+\frac{2\rho_2^2}{r^2}+ \frac{8k^{2}\rho_{1}\rho_{2}-m \rho_1^2 }{6 r^3} + \cdots~, \nonumber\\
W_0(r) \sim~ & -\frac{\rho_1^2}{2 r^2}- \frac{4\rho_1 \rho_2+3k\omega}{3r^3}+ \cdots~, \nonumber \\
W_1(r) \sim~ & \frac{k\omega}{ r^3}- \frac{k^{2}\rho_1^{2} }{2r^4}+ \cdots~, \nonumber \\
\rho( r)\sim~ & \frac{\rho_1}{r}+\frac{\rho_2}{r^2}+\frac{3k^{2}\rho_{1}-\rho_1^3 }{6r^3} +\cdots.
\end{align}
The behavior of these functions on the radial coordinate $r$ is plotted in Figure \ref{fig:HBH01} for a numerical solution. For more numerical results on this black branes solution, see \cite{Ahn:2019pqy}.
The temperature and the entropy density of this charged black brane are computed as
\begin{align}
T =\frac{1}{4\pi}e^{W_0(r_h)}U'(r_h)
 = \frac{r_h}{4\pi }e^{w_0}\Big(2+\cosh(2\rho_h)\Big)\,, \qquad
s = 2\pi r_h^2 e^{w_1}\,.
\end{align}

By the Noether method, the electric  charge per unit length of the two-form gauge field $C$ is given by
 \begin{equation}
 q\equiv \frac{1}{2}\int \sinh^{-2}(2\rho) \ast dC=\frac{1}{2}\int d\chi=\frac{1}{2}k~,
\end{equation}
which is nothing but  the number of branes stretching along the $y$-direction per unit length in the $x$-direction. This could be regarded as the `magnetic' charge of the phase scalar field $\chi$.
In this charge configuration,  the non-vanishing component of the two-form gauge field is given by
 \begin{equation}
C_{ty}(r)=k\int_{r_h}^{r}dr'\,e^{W_0-W_1}  \sinh^{2}(2\rho)~,
\end{equation}
that is chosen to vanish at the horizon to satisfy the regularity condition of the gauge field.
Then the associated chemical  potential of the charge density $q$ is defined as
  \begin{equation}
\Phi\equiv C_{ty}(\infty)=k\int_{r_h}^{\infty}dr\,e^{W_0-W_1}  \sinh^{2}(2\rho)=6\omega \,.
\end{equation}
The asymptotic expansion of $C_{ty}$ is given by
  \begin{align}
C_{ty}(r)&=k\int_{r_h}^{\infty}dr'\,e^{W_0-W_1}  \sinh^{2}(2\rho) -k\int_{r}^{\infty}dr'\,e^{W_0-W_1}  \sinh^{2}(2\rho) \nonumber\\
&=\Phi -\frac{4k\rho_1^2}{r} +\cdots~,
\end{align}
where the leading term corresponds to the chemical potential while the coefficient of the next-to-leading term does to the charge.

\section{Bulk Conserved Charges and Reduced Action Formalism}

In this section, we establish the black brane thermodynamics by finding bulk charges which satisfy the thermodynamic first law. The infinitesimal expressions of bulk charges can be integrated with proper boundary condition.  We also obtain the relation which connects the parameters at the horizon and those at the asymptotic boundary by using the scaling symmetry in the reduced action formalism.

\subsection{Bulk conserved charges}

We construct the well-defined conserved charge densities of the black brane, which play the role as thermodynamic variables  and satisfy  the first law of black hole thermodynamics.  In this paper, we use the off-shell ADT formalism developed in \cite{Kim:2013qra,Kim:2013cor,Hyun:2014kfa,Hyun:2014sha,Hyun:2014nma} to obtain these conserved charges.

Let us start with the brief review of the off-shell ADT formalism we use. In the original ADT formulation \cite{Abbott:1981ff,Deser:2002rt}, total conserved charges, like mass and angular momentum, are constructed via linearization of the gravitational field equation denoted by $\delta\mathcal{E}_g$, and the corresponding current $$\mathcal{J}^{\mu}=\delta\mathcal{E}_g^{\mu\nu}\xi_\nu\,,$$ is introduced for isometry generated by a Killing vector $\xi$. By using the fact that this current is conserved for the on-shell configuration, one can get the total charge expression which is given by the integration at the asymptotic boundary. 
As is well known, the entropy of the black hole can be interpreted as the Noether charge at the Killing horizon \cite{Wald:1993nt,Iyer:1994ys}. One may introduce quasi-local charges and extend the ADT formalism to incorporate the conserved charge at the event horizon as well as the conserved charges defined at the asymptotic infinity.

In order to achieve this extension, we promote the current, $\mathcal{J}^\mu$, to identically conserved current satisfying 
$\nabla_{\mu} \mathcal{J}^{\mu}_{ADT}=0$,
which does not need equations of motion. This current is constructed by adding appropriate terms, which will be clarified soon. This allows us to have the off-shell ADT potential, $Q^{\mu\nu}_{ADT}$, which is defined as $\mathcal{J}^{\mu}_{ADT}\equiv\nabla_{\nu}Q^{\mu\nu}_{ADT}$. We would like to note that the  potential is related to the linearized Noether potential and the surface terms introduced in  Wald's formulae in \cite{Wald:1993nt,Iyer:1994ys}. Therefore, we can obtain the conserved charges in a pure gravity theory by integrating the off-shell ADT potential on a codimension-2 hypersurface including the Killing horizon of the black holes.

Now we follow the procedure described in  \cite{Hyun:2014sha} where quasi-local conserved charges are constructed in the gravity theory coupled to matter fields. 
The generic variation of the action (\ref{bulk_action3}) with respect to the metric and matter fields can be expressed as
\begin{align}
        \delta S =\frac{1}{2} \int d^{4}x\,\sqrt{-g} \,\left( \mathcal{E}^{g}_{\mu\nu}\delta g^{\mu\nu} +\mathcal{E}_{\rho}\delta\rho +\mathcal{E}^{\mu\nu}_{C}\delta C_{\mu\nu}  + \partial_{\mu} \mathbf{\Theta}^{\mu}(\delta g, \delta \rho, \delta C)\right)\,,
\end{align}
where $\mathcal{E}_{\mu\nu}^{g},\mathcal{E}_{\rho}$ and $\mathcal{E}^{\mu\nu}_{C}$ denote the expressions of the field equations and $\mathbf{\Theta}^{\mu}$ stands for the surface term. The explicit expressions of the Einstein equation and the energy-momentum tensors are given by
\begin{align}
        \mathcal{E}_{\mu\nu}^{g} &\equiv G_{\mu\nu}-T_{\mu\nu}^{\rho}-T_{\mu\nu}^{C}\,, \nn\\
        G_{\mu\nu} &= R_{\mu\nu}-\frac{1}{2}g_{\mu\nu}R\,, \nn\\
        T^{\rho}_{\mu\nu} &= 2\partial_{\mu}\rho\, \partial_{\nu}\rho +\frac{1}{2}g_{\mu\nu}\left[-2(\partial\rho)^{2} +2(3+2\sinh^{2}\rho)\right]\,, \\
        T^{C}_{\mu\nu} &= \frac{1}{4}\sinh^{-2}(2\rho)F^{(3)}_{\mu\alpha\beta}F^{(3)}_{\nu}\,^{\alpha\beta} +\frac{1}{2}g_{\mu\nu}\left[-\frac{1}{12}\sinh^{-2}(2\rho)\,F_{(3)}^{\alpha\beta\gamma}F_{\alpha\beta \gamma}^{(3)}\right]\,,\nn
\end{align}
and the field equations for $\rho$ and $C$ are determined as
\begin{align}
        \mathcal{E}_\rho &= 4\nabla^2 \rho +2\coth(2\rho)\sinh^{-2}(2\rho)F_{(3)}^2 +4\sinh(2\rho)\,, \nn\\
         \mathcal{E}^{\mu\nu}_C &= \frac{1}{2} \nabla_\rho \left[\sinh^{-2}(2\rho)F^{\rho\mu\nu} \right]\,.
\end{align}
The surface term for each field variation can be written  as follows:
\begin{align}
        \mathbf{\Theta}^{\mu}(\delta g, \delta \rho, \delta C) &\equiv \mathbf{\Theta}_g^{\mu}(\delta g)+\mathbf{\Theta}^{\mu}_\rho(\delta \rho)+\mathbf{\Theta}^{\mu}_C(\delta C) \,,\nn\\[5pt]
        \mathbf{\Theta}^{\mu}_g(\delta g) &=\sqrt{-g}\, \left[ g^{\mu\nu}\nabla^{\lambda}\delta g_{\nu\lambda} -g^{\nu\lambda}\nabla^{\mu}\delta g_{\nu\lambda} \right]\,, \nn \\[5pt]
        \mathbf{\Theta}^{\mu}_\rho(\delta \rho) &= -4\sqrt{-g}\,\partial^{\mu}\rho \,\delta\rho \,,\\ \mathbf{\Theta}^{\mu}_C(\delta C) &= -\frac{1}{2} \sqrt{-g}\,\sinh^{-2}(2\rho)F_{(3)}^{\mu\alpha\beta}\delta C_{\alpha\beta}\,.\nn
\end{align}

Under the diffeomorphism  generated by the vector field $\zeta$, the following off-shell identity holds:
\begin{align}\label{off-shell_identity1}
        &2\zeta_{\nu}\nabla_{\mu} \mathcal{E}_g^{\mu\nu} +\mathcal{E}_\rho\pounds_\zeta \rho +\mathcal{E}_C^{\mu\nu}\pounds_\zeta C_{\mu\nu}=\nabla_{\mu}(\mathcal{Z}^{\mu\nu}\zeta_{\nu})\,,
\end{align}
where $\pounds_{\zeta}$ denotes the Lie derivative along $\zeta$ direction. 
The rank-2 tensor $\mathcal{Z}^{\mu\nu}$ comes from the matter fields due to the Bianchi identity. With a little algebra, one can show that $\mathcal{Z}^{\mu\nu}$ is proportional to the field equations of the matter fields and the only contribution comes from the the two-form gauge field as follows:
\begin{align}
        \mathcal{Z}^{\mu\nu}=2\mathcal{E}^{\mu\rho}_{C}\,C^{\nu}\,_{\rho} = -\nabla_{\alpha}\big(\sinh^{-2}(2\rho)F^{\mu\alpha\beta}_{(3)}\big) C_{\beta}\,^{\nu}\,.
\end{align}
This appears in the following construction of the off-shell ADT current.
%
%
Using the identity (\ref{off-shell_identity1}), the following equality holds:
\begin{align}\label{equality}
        [\pounds_{\zeta},\delta\, ](\sqrt{-g}\mathcal{L} ) &=\partial_{\mu}\bigg[ \delta\Big( \sqrt{-g}\big[\mathcal{E}^{\mu\nu}_{g}-\frac{1}{2}\mathcal{Z}^{\mu\nu} \big] \zeta_{\nu}\Big) +\frac{1}{2}\omega^{\mu}(\pounds_{\zeta}\psi,\delta\psi\, ) \nonumber\\
        &\qquad -\frac{1}{2}\sqrt{-g}\zeta^{\mu}\mathcal{E}^{\alpha\beta}_{g}\delta g_{\alpha\beta} +\frac{1}{2}\sqrt{-g}\zeta^{\mu}\mathcal{E}_\rho \delta\rho +\frac{1}{2}\sqrt{-g}\zeta^{\mu}\mathcal{E}^{\alpha\beta}_{C}\delta C_{\alpha\beta} \bigg]\,,
\end{align}
where $\mathcal{L}$ is the Lagrangian density of our model and $\psi=(g_{\mu\nu},\rho,C)$. The symplectic current $\omega^{\mu}$  is defined as
\begin{align}
        \omega^{\mu}(\delta_1 \psi,\delta_2 \psi) \equiv \delta_1\mathbf{\Theta}^{\mu}(\delta_2 \psi)-\delta_2\mathbf{\Theta}^{\mu}(\delta_1 \psi)\,.
\end{align}

The left-hand side of the Eq. (\ref{equality}) as well as the symplectic current  vanish when the vector $\zeta$ becomes Killing vector $\xi$. Naturally we can construct the ADT current $\mathcal{J}_{ADT}$ for our model by
\begin{align}\label{ADTcurrent}
        \sqrt{-g}\mathcal{J}^{\mu}_{ADT}(\xi) &\equiv \delta\Big( \sqrt{-g}\big[\mathcal{E}^{\mu\nu}_{g}-\frac{1}{2}\mathcal{Z}^{\mu\nu} \big] \xi_{\nu}\Big) \nonumber\\
         &\quad -\frac{1}{2}\sqrt{-g}\xi^{\mu}\mathcal{E}^{\alpha\beta}_{g}\delta g_{\alpha\beta} +\frac{1}{2}\sqrt{-g}\xi^{\mu}\mathcal{E}_\rho \delta\rho +\frac{1}{2}\sqrt{-g}\xi^{\mu}\mathcal{E}^{\alpha\beta}_{C}\delta C_{\alpha\beta}\,,
\end{align}
which is conserved identically. Consequently, we obtain the off-shell ADT potential from the divergence relation $\mathcal{J}^{\mu\nu}_{ADT}=\nabla_{\nu}Q^{\mu\nu}_{ADT}$.
In our case, the off-shell ADT potential can be expressed separately by the contributions of each field as
\begin{align}\label{ADTQ2}
	Q^{\mu\nu}_{ADT}(\xi;\, \delta g,\delta \rho,\delta C) =Q^{\mu\nu}_{ADT}(\xi;\, \delta g)+Q^{\mu\nu}_{ADT}(\xi;\, \delta \rho) +Q^{\mu\nu}_{ADT}(\xi;\, \delta C)\,.
\end{align}
Here we display explicit formulae for each term:
\begin{align*}
	Q^{\mu\nu}_{ADT}(\xi;\, \delta g) &= \xi^{[\mu}\nabla_{\alpha}\delta g^{\nu]\alpha} -\xi_{\alpha}\nabla^{[\mu}\delta g^{\nu]\alpha} -g_{\alpha\beta}\xi^{[\mu}\nabla^{\nu]}\delta g^{\alpha\beta} +\delta g^{\alpha[\mu}\nabla_{\alpha}\xi^{\nu]} \,,\nonumber \\
	&\quad -\frac{1}{2}g_{\alpha\beta}\delta g^{\alpha\beta} \Big(\nabla^{[\mu}\xi^{\nu]} -\frac{1}{2}\sinh^{-2}(2\rho)F^{\mu\nu\sigma}C_{\sigma\lambda}\xi^{\lambda} \Big) \nonumber \\
	&\quad +\sinh^{-2}(2\rho)\Big( \delta g^{\alpha[\mu}F_{\alpha}{}^{\nu]\beta} +\frac{1}{2}F^{\mu\nu}{}_{\alpha}\delta g^{\alpha\beta} \Big)C_{\beta\gamma}\xi^{\gamma} \,, \\[5pt]
	Q^{\mu\nu}_{ADT}(\xi;\, \delta \rho) &= 2\,\delta\rho \Big( 2\xi^{[\mu}\partial^{\nu]}\rho+\coth(2\rho)\sinh^{-2}(2\rho)F^{\mu\nu\alpha}C_{\alpha\beta}\xi^{\beta} \Big) \,,\\[5pt]
	Q^{\mu\nu}_{ADT}(\xi;\, \delta C) &= \frac{1}{2}\sinh^{-2}(2\rho)\Big( \xi^{[\mu}F^{\nu]\alpha\beta}\delta C_{\alpha\beta} -F^{\mu\nu\alpha}\xi^{\beta}\delta C_{\alpha\beta} -g^{\mu\alpha}g^{\nu\beta} \delta (F_{\alpha\beta\gamma})C^{\gamma}{}_{\lambda}\xi^{\lambda} \Big)\,.
\end{align*}

The infinitesimal expression for the conserved charges, $\delta Q_{\Sigma}$, associated with the Killing vector $\xi$  in terms of the potential $Q^{\mu\nu}_{ADT}(\delta g,\delta \rho,\delta C)$ is given by
\begin{align}
        \delta Q_\Sigma=\int_{\Sigma} d^{D-2}x_{\mu\nu}\sqrt{-g}\,Q^{\mu\nu}_{ADT}(\delta g,\delta \rho,\delta C)\,,
\end{align}
where $\int_{\Sigma} d^{D-2}x_{\mu\nu}$ is the integration over  the codimension-2 hypersurface $\Sigma$.
After applying the ansatz (\ref{BBrane00}) to  (\ref{ADTQ2}), we obtain the expression of the infinitesimal charge density on $\Sigma$ for the timelike Killing vector $\xi=\partial_t$ as
\begin{align}
	\delta Q_\Sigma &=\int_\Sigma dxdy\, e^{W_0+W_1} \bigg[ \frac{r}{2}\left(-2-rW_1'\right)\delta U -\frac{3}{2}e^{-2W_0}\sinh^{-2}(2\rho)C_{ty}C_{ty}' \delta W_0 \nonumber \\
	&\qquad\qquad -\frac{1}{2} \Big( 2rU(1+rW_0'-rW_1') -r^2U' +e^{-2W_0}\sinh^{-2}(2\rho)C_{ty}' \Big)\delta W_1 \nonumber \\
	&\qquad\qquad -r^2 U\delta W_1' -2r^2 \Big( U\rho' +2e^{-W_0}\coth(2\rho)\sinh^{-2}(2\rho)C_{ty}'C_{ty} \Big)\delta \rho \nonumber \\
	&\qquad\qquad -\frac{1}{2} e^{-2W_0}\sinh^{-2}(2\rho)C_{ty}\delta C_{ty}' \bigg]\,.
\end{align}
At the horizon, $\Sigma=\mathcal{H}$,  it becomes
\begin{align}
        \delta Q_{\mathcal{H}}
        &=\frac{1}{2}e^{w_0+w_1} \,r_h^2\big(2\delta r_h+r_h\delta w_1\big) \Big(2+\cosh(2\rho_h)\Big) \nonumber \\[5pt]
        &=  \frac{e^{w_0}U(r_h)' }{4\pi}  \delta\left(2\pi r_h^2e^{w_1}\right) =T\delta s \,.
\end{align}
On the other hand, the infinitesimal conserved charge density at the asymptotic boundary gives
\begin{align}
        \delta Q_{\infty} (\delta g,\delta\rho)=\delta M &= \delta m +3\delta(k\omega) +2 \delta(\rho_1 \rho_2) +2\rho_2\delta\rho_1~,   \nonumber \\
        \delta Q_{\infty} (\delta C)=-\Phi\delta q &= -3 \, \omega\delta k\,.
\end{align}

Let us consider the codimension-1 hypersurface $\mathcal{C}$ with interior boundary $\mathcal{H}$ at the horizon and asymptotic boundary at $r\rightarrow\infty$.
 Then  the integration of the current $\mathcal{J}^{\mu}_{ADT}$ over this hypersurface $\mathcal{C}$ vanishes on-shell:
\begin{align}
        \int_{\mathcal{C}} d^{D-1}x_{\mu} \sqrt{-g} \mathcal{J}^{\mu}_{ADT}\Big|_{\mathcal{E}=0,\delta\mathcal{E}=0}=0\,,
\end{align}
which gives the following relation
\begin{align}
        \int_{\infty} d^{D-2}x_{\mu\nu}\sqrt{-g}\,Q^{\mu\nu}-\int_{\mathcal{H}} d^{D-2}x_{\mu\nu}\sqrt{-g}\,Q^{\mu\nu} =\delta Q_{\infty}-\delta Q_{\mathcal{H}}=0\,.
\end{align}
This establishes the thermodynamic first law of this charged black brane as
$$
dM=Tds+\Phi dq\,.$$

In order to guarantee the integrability of the mass density, we need to impose a relation between two normalizable modes $\rho_{1}$ and $\rho_{2}$  in asymptotic expansion (\ref{AsympExp}) of the scalar field. This relation, denoted as  $\rho_2=\frac{dW(\rho_1)}{d\rho_1}$ for some function $W$, is deeply related to the  deformation of the dual conformal field theory \cite{Witten:2001ua}.
With this boundary condition, the mass density can be integrated as
\begin{equation}
M = m+3k\omega +2 \rho_1 \rho_2 + 2W  \,.
        \label{MassExpression}
\end{equation}

\subsection{Reduced action formalism and scaling symmetry}
In the numerical solution of the black hole with scalar hair, it is usually not easy to connect the parameters at the horizon with those at the asymptotic boundary. It is  not straightforward to obtain the Smarr relation of the black hole either. In this respect, the
 reduced action formalism accompanied by `scaling symmetry technique' developed in \cite{Banados:2005hm,Hyun:2015tia,Ahn:2015uza,Ahn:2015shg,Hyun:2016isn,Hyun:2017nkb,Kim:2019lxb,Erices:2019onl} is highly useful.
 We start with the following reduced action:
\begin{align}
\mathcal{S}_B = \int dr d^3x\, L_{red}~,
\end{align}
where the reduced Lagrangian $L_{red}$ is obtained by plugging the ansatz (\ref{BBrane00}) into the bulk action (\ref{bulk_action}).  In our case, it is given by
\begin{align}
        L_{red} &= \frac{e^{W_0+W_1}}{2}\bigg[ r^2U'W_1'+2rUW_0'\big( 1+rW_1' \big) -2r^2U\rho'{}^2 \nonumber\\
        &\qquad\qquad\qquad +\frac{ 1}{2}e^{-2W_0}\sinh^{-2}(2\rho)(C_{ty}')^2 + r^2\big(5+\cosh(2\rho)\big) \bigg]\,,
\end{align}
 up to total derivative terms with respect to $r$.

Now we consider the scaling transformation of the reduced action. Generically a field $\Psi(r)$ with scaling weight $w_{\Psi}$  transforms as $\delta_\sigma \Psi= w_{\Psi }\Psi- r\Psi'$. If we assign weights of those functions $(U, e^{W_0}, e^{W_1}, \rho)$ as $(2, -2, -1, 0)$, then the variation of the  reduced Lagrangian is given by the total derivative as
\begin{equation}
        \delta_\sigma L_{red}=-(rL_{red})'\,,
        \label{varL1}
\end{equation}
up to the equations of motion of the reduced action. Thus the action is invariant under the scaling transformation.\footnote{
One may note that this choice of weights is not the unique choice but a convenient one.}  On the other hand, the general variation of the reduced Lagrangian is given by
\begin{equation}
        \delta L_{red}={\cal E}_{\Psi}\delta\Psi+\Theta'(\delta\Psi)\,,
        \label{varL2}
\end{equation}
where $\Psi$ denotes functions, $U,W_0,W_1$,  and $\rho$, collectively and $\mathcal{E}_\Psi$ is the corresponding equation of motion operator.  By equating the above two variations in Eqs. (\ref{varL1}) and Eqs.(\ref{varL2}) for the variation under scaling along with the on-shell condition $\left({\cal E}_{\Psi}=0\right)$, one obtains the relation:
\begin{equation}
        \big(\Theta(\delta_\sigma \Psi)+rL_{red}\big)'=0\,.
\end{equation}
 This tells us that the quantity,
\begin{align}
c(r)\equiv\Theta(\delta_\sigma \Psi)+rL_{red}\,,
\end{align}
 can be regarded  as  a conserved `charge' along the radial direction $r$.
The  charge function $c(r)$ can be expressed as
\begin{align}
c(r) = \frac{e^{W_0+W_1}}{2}\bigg[r^2U'+2rU(-1+rW_0')\bigg]\,.
\end{align}

Since the `charge' is conserved, i.e. $c'(r)=0$, it has the same value for any $r$, and in particular $c(r_{h})=c(r\rightarrow\infty)$.  At both locations, the charge function is related to the thermodynamic variables defined there  and this provides a nontrivial relation between those quantities. In our case at hand, the values of the charge function are given by
\begin{align}
        c(r_h) &= \frac{1}{2}r_h^2e^{w_0+w_1}U'(r_h) = s T~, \nn\\
        c(\infty) &= \frac{1}{2} \, \Big(3m +8 \rho_1 \rho_2 +6k\omega\Big)~.
\end{align}


\section{The Thermodynamics of the Neutral Black Brane and its Dual CFT}
In order to establish the the relation between the free energy and the on-shell action in the bulk as well as in the dual conformal field theory, we need to specify the function $W(\rho_1)$, implying the boundary condition of the scalar field, in the expression of mass density, (\ref{mass}). It is deeply related to the deformation of the dual CFT. In this section we describe the thermodynamics of the neutral black brane and the thermodynamics of the dual CFT. In particular, we consider the case corresponding to the marginal deformation by the conformal dimension three operator. In section 6, we consider the case of the deformation 
from other external sources. 

\subsection{The black brane thermodynamics}

The  two non-vanishing normalizable modes, $\rho_{1}$ and $\rho_{2}$, of the scalar field $\rho$ correspond to the source and the expectation value of the operator $\mathcal{O}$ with the conformal dimension $\Delta=2$ in the dual CFT, respectively.
In this case, the appropriate boundary condition of the scalar fields is given by \cite{Witten:2001ua}
\begin{equation}\label{relation1}
\rho_{2}= \frac{dW}{d\rho_{1}}=\nu \rho_{1}^{2}\,,\qquad  \qquad W(\rho_{1})=\frac{\nu}{3}\rho_{1}^{3}\,,
\end{equation}
or
\begin{equation}
\rho_{1}= \frac{d{\tilde W}}{d\rho_{2}}=\frac{1}{\sqrt{\nu}}\sqrt{ \rho_{2}}\,,\qquad  \qquad \tilde W(\rho_{2})=\frac{2}{3\sqrt{\nu}}\rho_{2}^{\frac{3}{2}}\,.
\end{equation}
In the dual conformal field theory,  this can be interpreted as the deformation by a marginal operator,
\begin{align}
        I_{CFT}\rightarrow I_{CFT}-\int d^3x\, \tilde W[{\cal O}]
        \label{deform1}
\end{align}
with the operator ${\cal O}$  dual to the scalar field $\rho$.
In this case, by plugging (\ref{relation1}) into (\ref{MassExpression}), the mass density of black brane is given by
\begin{equation}
        M=m   + \frac{8}{3}\rho_1 \rho_2 \,.
        \label{mass1}
\end{equation}
Furthermore, the charge function $c(r)$ gives the Smarr-like relation:
\begin{align}
        2sT &= 3M.\label{Smarr-like}
\end{align}

Now let us discuss the black brane thermodynamics in the extended phase space. If the cosmological constant can be taken as a dynamical variable, it would be considered as the pressure in the extended phase space as
\begin{align}
P\equiv -\Lambda=3\,,
\end{align}
and its conjugate thermodynamic volume density can be determined  as
\begin{align} \label{volume}
V\equiv \left( \frac{\partial M}{\partial P}\right)_{S}.
\end{align}
The thermodynamic volume obtained in this fashion usually turns out to be the so-called `volume' surrounded by the event horizon and the pressure can be regarded as the pressure on the surface of the event horizon. 
This gives the interpretation of $PV$ as the energy required to place the black hole in the environment with the pressure produced by the cosmological constant and the total energy $M$ as the enthalpy, which includes not only the internal energy but also the energy $PV$.

This suggests that, in the case of the AdS black hole with scalar hair,  it would be better to consider the radial pressure on the event horizon due to the scalar field. Therefore we define the thermodynamic pressure as
\begin{align}
        P\equiv T_r^{r}\Big|_{r_h}= 3+ 2\sinh^{2} \rho_h\,.
\end{align}
By using the relation
\begin{align}   
        (3+ 2\sinh^{2} \rho_h)=\frac{2}{r_h^3}e^{-w_0}sT
        = \frac{1}{r_h^3} e^{-w_0} \left(3 m   + 8\rho_1 \rho_2 \right)
        \label{rho0}
\end{align}
from the charge function $c(r)$, the  thermodynamic pressure can be computed as
\begin{align}   
        P&= \frac{1}{r_h^3}e^{-w_0}\left(3m  + 8\rho_1 \rho_2 \right)\,.
        \end{align}
Then the conjugate thermodynamic volume density, defined in (\ref{volume}), is  determined as
\begin{align}
V =\frac{r_h^3}{3} e^{w_0},
\end{align}
which, indeed, reminds the volume density of the black brane behind the horizon.
Using these quantities, the charge function $c(r)$ gives
 the Smarr relation in the extended phase space as
\begin{align}
        M &= 2Ts-2PV.
\end{align}

\subsection{The thermodynamics of the marginally deformed ABJM theory}

In order to describe the holographic renormalization in the dual CFT, we begin with the ADM decomposition of the metric along the radial direction:
\begin{align}\label{ADM}
ds^2 = N(r)^2 dr^2 + \gamma_{ij} dx^i dx^j\,
\end{align}
with the boundary coordinates $x^i$ and the boundary metric $\gamma_{ij}$. 
The total action is given by
\begin{align}
S_{\rm tot} = S_B +S_{\rm GH}+ S_{\rm ct} + S_{3}~,
\end{align}
where the bulk action $S_B$ is given in (\ref{bulk_action2}).  The Gibbons-Hawking term $S_{\rm GH}$ for the well-posed variational problem of the bulk metric is given by
\begin{align}\label{SGH}
S_{\rm GH} &=   \int_{\partial \mathcal M} d^3 x \sqrt{-\gamma}\,K \,,
\end{align}
where  the extrinsic   curvature is given by  $K_{ij}=\frac{1}{2N}\gamma'_{ij}$ and $K= \gamma^{ij}K_{ij}$.
The counter term $S_{\rm ct}$ for the cancellation of the divergencies coming from the bulk action is given by
\begin{align}
S_{ct} &=-2 \int d^3 x\sqrt{-\gamma} \left(1 +\frac{1}{2}\rho^2  \right)\,. \label{Sct}
\end{align}
In addition to these usual boundary terms, we need an additional boundary term $S_{3}$
 \begin{align}\label{S3}
 S_{3} &=-2 \int d^3 x \sqrt{-\gamma} \,\, W(\rho)
 =-2\int d^3 x \sqrt{-\gamma} \,\,\,\frac{\nu}{3}  \rho^3~,
\end{align}
to implement the deformation given in (\ref{deform1}).

With these boundary action terms, the Euclidean on-shell action per unit volume can be computed as
$$
\frac{1}{\beta}S_{on-shell}=F
= -\frac{1  }{2}  \left(m  + \frac{8}{3}\rho_1 \rho_2 \right)\,,
$$
 where $F$ is the Helmholtz free energy density in the dual CFT. 
The boundary energy momentum tensor defined by the variation of the total action with respect to the boundary metric $\gamma_{ij}$ is given by the expression:
\begin{align}
\left<T^{ij}\right>=\lim_{r\to \infty} r^5 \left(K \gamma^{ij}-K^{ij}  - \frac{1}{2} \gamma^{ij}\left( 4 +2\rho^2 +\frac{4\nu}{3}  \rho^3 \right) \right)\,.
\end{align}
Thus the energy density in the dual CFT can be evaluated as
$$\epsilon\equiv \langle T^{tt}\rangle= m + \frac{8}{3}\rho_1 \rho_2 \,,$$
which matches exactly with the bulk mass density $M$, consistent with the AdS/CFT correspondence. 
The pressure in the dual CFT can be computed as
$$P=\langle T^{xx}\rangle=\langle T^{yy}\rangle=\frac{1  }{2}  \left(m  + \frac{8}{3}\rho_1 \rho_2 \right)\,.$$
The boundary energy momentum tensor $T^{ij}$ is conserved and traceless and thus satisfies the Ward identities as expected.
It is also clear that the  thermodynamic relation in the dual conformal field theory is established as
\begin{align}\label{HFreeEnergy}
F = \epsilon - s T~.
\end{align}

\section{The Thermodynamics of the Charged Black Brane and its Dual CFT}
Now we turn to our main interests: the black brane geometry with modulation, or charged black brane geometry. If it is viewed as the black brane geometry with modulated scalar fields, the boundary CFT is the mass deformed ABJM model with spatial modulation in the mass. Alternatively, it could be considered as the charged black brane solution and the corresponding dual CFT would be the marginal deformation 
in the three-form field flux background. 

Apparently, the corresponding bulk geometries look identical, but actually they are different as they have different scalar profile and, as a result, different mass. The black brane with the scalar modulation has the scalar hair with  two normalized scalar modes related linearly.
Namely, the asymptotic expansion of the scalar field $\rho$ is given by 
\begin{align}\label{AsympExp1}
\rho( r)\sim~ & \frac{\rho_1}{r}+\frac{\alpha\rho_1}{r^2} +\cdots\,,
\end{align}
where $\alpha$ is proportionality constant.
On the other hand, the charged black brane has the scalar fields with two modes behaving in the same manner as  those in the neutral black brane, 
\begin{align}\label{AsympExp2}
\rho( r)\sim~ & \frac{\rho_1}{r}+\frac{\nu\rho_1^{2}}{r^2} +\cdots\,.
\end{align}
In the vanishing charge limit, $q\rightarrow 0$, it reduces to the neutral black brane described in the previous section. 
Both of these solutions with $k\neq 0$ can be realized numerically.

In this section, we present the thermodynamics of the charged black brane and its dual CFT with the boundary condition (\ref{AsympExp2}). In the next section we deal with the  black brane corresponding to  the spatially modulated scalar field with the boundary condition (\ref{AsympExp1}).
 We begin with the description of the black brane thermodynamics.

\subsection{The charged black brane thermodynamics}
Since we are dealing with the charged black brane solution as the dual geometry of the CFT with marginal deformation (\ref{deform1}), we use the same relation (\ref{relation1}) between two scalar modes, $\rho_{1}$ and $\rho_{2}$,  as the one in the neutral black brane. 
From the general expression in Eqs. (\ref{MassExpression}), the mass density of the black brane becomes
\begin{equation}
        M=m + \frac{8}{3}\rho_1 \rho_2 +3k\omega \,.
        \label{mass}
\end{equation}

In parallel with the discussion of the neutral black brane, we would like to describe the black brane thermodynamics in the extended phase space. 
We introduce the radial pressure on the surface of the event horizon as 
\begin{align}
        P= P_{0}+P_{1}= (2+\cosh(2\rho_{h}))- \frac{k^2 e^{-2w_1}\sinh^2{2\rho_{h}}}{4r_{h}^2}\,,
\end{align}
where the pressure $P_{0}$ comes from the scalar potential, while the pressure $P_{1}$ does from the two-form gauge field. The  contribution of  this radiation pressure  $P_{1}$ would be much smaller than the scalar pressure $P_{0}$  and negligible.  This can also be understood as  follows.
The black brane geometry has a Killing vector $\xi^x =\frac{\partial}{\partial x}$, which means that the distribution of the two-form field charge or the modulation of the scalar profile in the $x$-direction does not affect the geometry much.
The existence of the charge distribution only gives rise to the slight change of the pressure. 

By using the relation from the charge function $c(r)$, the pressure  can be expressed as 
\begin{align}
        P\simeq P_{0}= \frac{e^{-w_0-w_1}}{ r_h^3}(3m +8\rho_1\rho_2+6k \omega) \,,\label{pressure}
\end{align}
and the conjugate thermodynamic volume density is given by
\begin{align}
V\equiv \left( \frac{\partial M}{\partial P}\right)_{S, k}\simeq\frac{r_h^3}{3} e^{w_0+w_{1}}\frac{m + \frac{8}{3}\rho_1 \rho_2 +3k\omega}{m + \frac{8}{3}\rho_1 \rho_2 +2k\omega}     \,.\nn
\end{align}
By plugging these expressions in the charge function relation, 
 we establish the Smarr relation for the charged black brane as
\begin{align}
        M &= 2Ts-2PV+\Phi q~.
\end{align}

\subsection{The thermodynamics of the marginally deformed ABJM theory}
In order to obtain the on-shell action and the boundary energy-momentum tensor,  we use the same boundary action terms as those in the case of neutral black brane.
The boundary energy density is found to be
\begin{align}
\epsilon\equiv  \langle T^{tt} \rangle = m +\frac{8}{3} \rho_1 \rho_2+3k\omega\,,
\end{align}
which matches exactly with the bulk mass density $M$ in Eq. (\ref{mass}).
The pressure in the $x$- and $y$-directions are computed as
\begin{align}
P_{x} &\equiv  \langle T^{xx} \rangle = \frac{1}{2} \, \Big(m  +\frac{8}{3} \rho_1 \rho_2 +6k\omega  \Big)\,, \nonumber \\
P_{y} &\equiv  \langle T^{yy} \rangle = \frac{1}{2} \, \Big(m  +\frac{8}{3} \rho_1 \rho_2 \Big)\,. 
\end{align}
The pressure anisotropy, $\Delta P\equiv P_{x}-P_{y}$, comes from  the existence of periodically spaced branes in the $x$-direction and the corresponding chemical potential,\footnote{One nice example of the pressure anisotropy is the anisotropic plasma which is described in \cite{Mateos:2011tv}, in detail. In that system, the pressure anisotropy arises due to the dissolved D7-brane and the resultant chemical potential as well.
}
\begin{align}
\Delta P= q\,\Phi= 3 k\omega\, \,.\label{qPhi}
\end{align}

With the given boundary action terms, we are working in the canonical ensemble with the constant temperature and charge density, and therefore the Euclidean on-shell action per unit volume is related to the Helmholtz free energy  and is determined as
$$
\frac{1}{\beta}S_{on-shell}=F
= -\frac{1}{2} \, \Big(m +\frac{8}{3} \rho_1 \rho_2 \Big).
$$
One may note that it is related to the  pressure $P_y$ in the $y$-direction as will be clarified in below.
By using the relation in the charge function $c(r)$, the Helmholtz free energy density $F(T, q)$ in the canonical ensemble satisfies the
 thermodynamic relation in the boundary CFT as
\begin{align}\label{HFreeEnergy1}
F = \epsilon - s T=-P_{y}\,.
\end{align}
From the thermodynamic first law, 
\begin{align}
d \epsilon =  T ds +\Phi dq\,,
\end{align}
the free energy density also satisfies
\begin{align}
dF =  - s dT + \Phi dq\,,
\end{align}
as expected.

The thermodynamic potential in the grand canonical ensemble, $G(T, \Phi)$, which would have replaced the Helmholtz free energy in relation with the on-shell action  if we  included additional boundary action term on the two-form gauge field, is given by
\begin{align}
G=\epsilon-Ts -\Phi q=  -\frac{1}{2} \, \Big(m  +\frac{8}{3} \rho_1 \rho_2 +6k\omega  \Big)\,.
\end{align}
The thermodynamic potential  is related to the  pressure $P_x$ in the $x$-direction as $G=-P_{x}$ and satisfies the relation
\begin{align}
dG=-s dT - q d\Phi\,.
\end{align}

Now we include volume $V=l_{x}l_{y}$ of the box and define  the total energy $E=\epsilon V$, total entropy $S=sV$, total charge $Q=ql_{x}$, and potential $\Phi_{C}=\Phi l_{y}$.
Then the first law takes 
the form
\begin{align}
dE= T dS + \Phi_{C} dQ-l_{x}P_{y}dl_{y}-l_{y} P_{x}dl_{x}\,,
\end{align}
where $l_{x}P_{y}$, $l_{y}P_{x}$ are generalized forces corresponding to the displacement $dl_{y}$, $dl_{x}$, respectively, and are given by
\begin{align}
l_{x}P_{y}=-\left(\frac{\partial E}{\partial l_{y}}\right)_{S, l_{x}, Q},\qquad  l_{y}P_{x}=-\left(\frac{\partial E}{\partial l_{x}}\right)_{S, l_{y}, Q}~,
\end{align}
while the temperature and the potential satisfy
\begin{align}
T=\left(\frac{\partial E}{\partial S}\right)_{l_{x},l_{y}, Q}, \qquad \Phi_{C}=\left(\frac{\partial E}{\partial Q}\right)_{S,l_{x},l_{y}}. 
\end{align}
The total energy in the box $E(S, Q, l_{x}, l_{y})$ is an extensive quantity satisfying the scaling relations
 \begin{align}
 E(aS, aQ, al_{x}, l_{y})&=   aE(S, Q, l_{x}, l_{y})\,,\nn\\
  E(aS, Q, l_{x}, al_{y})&=   aE(S, Q, l_{x}, l_{y})\,. 
\end{align}
By using  the Euler's theorem on the function with scaling laws,\footnote{The Euler's theorem states that if a function $f(\omega_{1},\cdots, \omega_{n})$ obeys the scaling relation $f(a^{\alpha_{1}}\omega_{1}, \cdots a^{\alpha_{n}}\omega_{n})=a^{p}f(\omega_{1},\cdots, \omega_{n})$, then the function satisfies
\[
pf(\omega_{1},\cdots, \omega_{n})=\sum_{i=1}^{n}\alpha_{i}\left(\frac{\partial f}{\partial\omega_{i}}\right)\omega_{i}~.
\]
} 
we establish the thermodynamic relation in the dual CFT as
 \begin{align}
E= T S + \Phi_{C} Q_{C}-P_{x}V= T S -P_{y}V\,.
\end{align}
This also confirms the relations, $F=-P_{y}$ and $G=-P_{x}$.

\section{The Thermodynamics of the  Black Brane with Modulated Scalar Hair and its Dual ABJM Model}
In this section, we present the thermodynamics of the  black brane with the spatially modulated scalar field. The black brane geometry is believed to be  dual to the mass deformed ABJM model with spatial modulation.
In order to describe it as the dual geometry of the ImABJM model, we should go back to the original action with the complex scalar field $z$. The real and imaginary parts of the field $z=X+i\, Y$ have different origin from the 11-dimensional supergravity as scalar and psedo-scalar, respectively. Furthermore their dual boundary operators, $\mathcal{O}_X$ and  $\mathcal{O}_Y$, in the massive ABJM model are given by\footnote{Here $Y$'s and $\psi$'s denote the scalar field and the fermion field in the ABJM model and $M_A^B$ is given by $\text{diag}(1,1,-1,-1)$.}
\begin{align}\label{SusyDeformation}
\left\{\mathcal{O}_{ X},\mathcal{O}_{ Y} \right\} = \left\{ M_A^B  \tr \left(Y^A Y_B^\dagger\right) ,\, M_A^B \tr\left( \psi^{\dagger A}\psi_B + \frac{8\pi}{q} Y^C Y^\dagger_{[C} Y^A Y^\dagger_{B]} \right) \right\}
\end{align}
with the conformal dimension $\Delta_X=1$ and  $\Delta_Y=2$, respectively.  This means that the role of the two asymptotic modes for the field $X$ should be interchanged by the Legendre transformation~\cite{Klebanov:1999tb}.

\subsection{The thermodynamics in dual ABJM model}

In  the holographic renormalization with the ADM decomposition of the metric (\ref{ADM}),
the momentum conjugate to the scalar field $X$ is given by
\begin{align}
\pi_X & =- 4\sqrt{-\gamma}\,  n^\mu\,\partial_\mu X \,,
\end{align}
where $n^\mu$ denotes the normalized vector along the radial direction. 
The counter term (\ref{Sct}) not only cancels the divergence but also shifts the conjugate momentum of the scalar field $X$ as
\begin{align}
 \Pi_X &= \pi_X + \frac{\delta S_{ct}}{\delta X} =-4 \sqrt{-\gamma}\, \left( n^\mu\,\partial_\mu X+X\right)\,.
\end{align}

We would like to have the operator $\mathcal{O}_X$ dual to the scalar field $X$ with conformal dimension $\Delta_X=1$. This means that the role of two normalizable modes of the scalar field are interchanged, namely, $\rho_{1}$ and  $\rho_{2}$ become the expectation value and the source of the dual operator   $\mathcal{O}_X$, respectively. This  can be achieved by performing the Legendre transformation with respect to the scalar field $X$. This means that we should take the boundary term $S_2$, instead of $S_3$ in (\ref{S3}), which is given by
\begin{align}
S_{2} &=\frac{1}{2} \int d^3 x \sqrt{-\gamma}   J_X\,X\,.
\end{align}
Here the source $J_{X}$ for the operator $\mathcal{O}_X$ is defined by
\begin{align}
J_X & \equiv - \frac{1}{\sqrt{-\gamma} }\Pi_X\,,
\end{align}
and is determined as
\begin{align}
J_X & =4 \, \left( n^\mu\,\partial_\mu X +X\right)\,.
\end{align}
Hence the additional boundary term becomes
\begin{align}
S_{2} &=2 \int d^3 x \sqrt{-\gamma}  \left( X\, n^\mu\partial_\mu X + X^2 \right)\,.
\end{align}

The vacuum expectation values of the chiral primary operators in the dual ABJM model are determined as\footnote{Vacuum expectation values of ${\cal O}_X$ and ${\cal O}_Y$ of the ABJM theory with a constant mass deformation in the large $N$ limit were obtained~\cite{Jang:2016aug, Jang:2019pve} in terms of the Kaluza-Klein holography method.}
\begin{align}
\left<\mathcal{O}_X\right>& =\lim_{r\to \infty} \frac{1}{2r^{2}} \sqrt{-\gamma} X  ~
=\frac{\rho_1\,}{2}\,  \cos  {k}  {x}~,
\label{vevX}\\
\left<\mathcal{O}_Y\right> &=\lim_{r\to \infty}\frac{2}{r }  \left(- \sqrt{-g}\frac{1}{1-|z|^2} \nabla^r Y - \sqrt{-\gamma}Y \right)
= 2\rho_2  \,   \sin  {k} {x}~,\label{vevY}
\end{align}
for their sources given by
\begin{align}
J_X = -4\rho_2   \cos {k}{x}~,\qquad J_Y = \rho_1  \sin {k}{x}~.
\end{align}
One may note that the neutral black brane with the boundary conditions described in this section can be found with the limit $k\rightarrow 0$. 

From the expression of the boundary energy momentum tensor, 
\begin{align}
\left<T^{ij}\right>=\lim_{r\to \infty}  r^5 \left( K \gamma^{ij} -K^{ij}  - \gamma^{ij}\left( 2 +Y^2 -X^2-2X\, r\, \partial_r X  \right) \right)\,,
\end{align}
the average boundary energy density is evaluated as
\begin{align}
\epsilon\equiv  \overline{\langle T^{tt} \rangle }= m +3k\omega +3 \rho_1 \rho_2\,,
\end{align}
and the pressure in the $x$- and $y$-directions are computed as
\begin{align}
P_{x} &\equiv  \overline{\langle T^{xx} \rangle } = \frac{1}{2} \, \Big(m  +6k\omega +2 \rho_1 \rho_2 \Big)\,, \nonumber \\
P_{y} &\equiv  \overline{\langle T^{yy} \rangle } = \frac{1}{2} \, \Big(m + 2 \rho_1 \rho_2 \Big)\,. 
\end{align}
The boundary energy momentum tensor satisfies two Ward identities
\begin{align}
\partial_i \left< T^{ij} \right>  =&  \left<\mathcal{O}_X\right> \partial^j J_{X}+  \left<\mathcal{O}_Y\right> \partial^j J_{Y}~, \\
\left< T^i_i \right> =& \left(3-\Delta_X\right)\left<\mathcal{O}_X\right>   J_{X}+\left(3-\Delta_{Y}\right)  \left<\mathcal{O}_{Y}\right>   J_{Y}~,
\end{align}
where the conformal anomaly is driven by the source. 
The Euclidean on-shell action per unit volume is evaluated  as
$$
\frac{1}{\beta}S_{on-shell}=F
= -\frac{1}{2} \, \Big( m+2\rho_1 \rho_2 \Big),
$$
and  the thermodynamic relation in the dual mass deformed ABJM model is established as
\begin{align}\label{HFreeEnergy2}
F = \epsilon - s T\,.
\end{align}
All the other thermodynamic relations given in section 5.2 follow accordingly.

\subsection{The black brane thermodynamics}
As the geometry is dual to the spatially varying mass deformation of the ABJM model,  the boundary condition for the scalar fields should be given by the linear relation between two normalizable modes as
        $$\rho_{2}=\mu\,  \rho_{1}\,\qquad {\rm or} \qquad W(\rho_{1})=\frac{\mu}{2}\rho_{1}^{2}~.$$
In this case  the mass of the black brane can be determined from the general expression in Eqs. (\ref{MassExpression}) as
\begin{equation}
        M=m +3k\omega +3 \rho_1 \rho_2  \,,
        \label{mass2}
\end{equation}
which agrees with the boundary energy density.
In the extended phase space,  the pressure  is given by the same expression given in (\ref{pressure})   
and the conjugate thermodynamic volume density is determined as
\begin{align}
V\equiv \left( \frac{\partial M}{\partial P}\right)_{S, k}\simeq\frac{r_h^3}{3} e^{w_0+w_{1}}\frac{m+ 3\rho_1 \rho_2 +3k\omega}{m + \frac{8}{3}\rho_1 \rho_2 +2k\omega}     \,.\nn
\end{align}
Therefore we obtain the same Smarr relation for this black brane:
\begin{align}
        M &= 2sT-2PV+\Phi q~.
\end{align}


\section{Discussion}

The Q-lattice black brane solution in 4-dimensions, which was introduced  in our previous work~\cite{Ahn:2019pqy} as a numerical solution, is dual to the ImABJM model with the sinusoidal mass function along a spatial direction $x$ at a finite temperature. In this paper, we also considered a charged black brane for the same spatial modulation by introducing the two-form gauge field. Then we constructed consistent thermodynamic relations for the Q-lattice  black brane and the charged black brane in the bulk gravity and the dual CFT simultaneously. 

To give a concrete physical interpretation for the spatial modulation parametrized by $k$, we started from an alternative form of the action by the field redefinition of the complex scalar field, $z = \tanh \rho\, e^{i\chi}$. Then we also introduced a two-form gauge field $C_{\mu\nu}$ whose field strength is Poincar\'e dual to the phase scalar field $\chi$. Turing on the phase scalar field as $\chi = kx$, one can give the charge $q= \frac{1}{2}k$  to the dual two-form gauge field $C_{\mu\nu}$. The presence of the charge denotes a periodic distribution along  the $x$-direction for charged branes spanning in the $y$-direction, and so it induces the anisotropy of thermodynamic variables  in the $(x,y)$-plane. The brane configuration for the charge in 11-dimensional supergravity point of view would be the periodic array of M5-branes along a worldvolume direction of M2-branes, as we argued previously. Under this circumstance, we successfully constructed the thermodynamic first laws and the Smarr relations  in the bulk gravity and various thermodynamic relations in the boundary dual CFT simultaneously  by using various methods, such as the off-shell ADT formalism, reduced action formalism, and holographic renormalization method.

According to the relation between coefficients, $\rho_1$ and $\rho_2$, in the asymptotic expansion of the scalar field $\rho$,    thermodynamic variables in the bulk gravity  and those in the  corresponding deformed  CFTs can be changed, even though in all cases  the same thermodynamic relations hold.  
In this paper, we analyzed two cases, $\rho_2 =\nu \rho_1^2$ and $\rho_2 =\mu \rho_1$. The former case corresponds to the charged (or neutral) black brane solution with   the charge $q$ and the chemical potential $\Phi$, which is  defined by the asymptotic boundary value of $C_{ty}$.  We obtained the first law and the mass expression of the black brane,  and constructed the Smarr relation by defining the  pressure as the $T^r_r$-component and the conjugate thermodynamic volume at the horizon in the extended phase space.   Then the  dual CFT is identified as  a marginal deformation by the gauge invariant operator with conformal dimension $\Delta= 3$. We obtained the Euclidean on-shell action, which is proportional to the Helmholtz free energy,  and the boundary energy-momentum tensor from the holographically renormalized boundary action. The Helmholtz free energy density in the canonical ensemble satisfies the well-defined thermodynamic relation in the boundary CFT. We also discussed the thermodynamic relation with the anisotropic pressure on the boundary CFT by defining the thermodynamic potential including the chemical potential term in the grand canonical ensemble. 

On the other hand,  for the Q-lattice black brane, which  is dual to the ImABJM model with the sinusoidal mass function, the boundary condition of the scalar  field should be the latter case, $\rho_2 =\mu \rho_1$.  Using the similar method with the case of the charged black brane, we obtained the consistent thermodynamic relations in the bulk gravity and in the dual ABJM theory.

In this paper, we constructed the thermodynamics of several representative hairly AdS black branes with anisotropic pressures induced from the modulated mass function. To do that, we employed the energy-momentum tensors and charges,  and their conjugate variables in the bulk gravity and the boundary CFT. It would be very interesting to generalize our construction for more diverse anisotropic cases. For instance, one can consider the modulation along the $y$-direction additionally and also extend to the higher dimensional black holes with anisotropy. Investigating the zero temperature limit of the charged black brane in more general setup is also interesting . We leave the problem as a future work.

\section*{Acknowledgments}

This work was supported by the National Research Foundation of Korea(NRF) grant with grant number NRF-2016R1D1A1A09917598 (B.A., S.H., S.P.), NRF-2019R1A2C1007396 (K.K.), and    NRF-2017R1D1A1A09000951, NRF-2019R1F1A1059220, NRF-2019R1A6A1A10073079 (O.K.). K.K. acknowledges the hospitality at APCTP where part of this work was done. S.P. acknowledges the hospitality at KIAS where part of this work was done.


\end{document}